# Thermoelectric Properties of Antiperovskite Calcium Oxides $Ca_3PbO$ and $Ca_3SnO$

Yoshihiko Okamoto,[1,2,a)] Akira Sakamaki,[1)] and Koshi Takenaka[1)]

[1]*Department of Applied Physics, Nagoya University, Nagoya 464-8603, Japan*

[2]*Institute for Advanced Research, Nagoya University, Nagoya 464-8601, Japan*

We report the thermoelectric properties of polycrystalline samples of $Ca_3Pb_{1-x}Bi_xO$ ($x$ = 0, 0.1, 0.2) and $Ca_3SnO$, both crystallizing in a cubic antiperovskite-type structure. The $Ca_3SnO$ sample shows metallic resistivity and its thermoelectric power approaches 100 μV $K^{-1}$ at room temperature, resulting in the thermoelectric power factor of $Ca_3SnO$ being larger than that of $Ca_3Pb_{1-x}Bi_xO$. On the basis of Hall and Sommerfeld coefficients, the $Ca_3SnO$ sample is found to be a *p*-type metal with a carrier density of ~$10^{19}$ $cm^{-3}$, a mobility of ~80 $cm^2$ $V^{-1}$ $s^{-1}$, both comparable to those in degenerated semiconductors, and a moderately large hole carrier effective mass. The coexistence of moderately high mobility and large effective mass observed in $Ca_3SnO$, as well as possible emergence of a mutivalley electronic structure with a small band gap at low-symmetry points in *k*-space, suggests that the antiperovskite Ca oxides have strong potential as a thermoelectric material.

## I. INTRODUCTION

Thermoelectric energy conversion is a solid-state conversion between thermal and electric energies by conduction electrons and is expected to be widely applied to environmentally-friendly power generators and refrigerators. The efficiency of thermoelectric conversion depends on the dimensionless figure of merit, $ZT = S^2T/\rho\kappa$, of a thermoelectric material, where $S$, $\rho$, $\kappa$, and $T$ are thermoelectric power, electrical resistivity, thermal conductivity, and operating temperature, respectively. Since these properties cannot be manipulated independently, the search for improved materials with larger $ZT$ is always challenging. One of the most challenging tasks is the enhancement of $S$, while maintaining low $\rho$. Various concepts for enhancing $S$, such as multi-valley electronic structures,[1] low dimensionality,[2,3] pudding-mold-type bands,[4,5] band convergence,[6,7] and strong electron correlation effects,[8,9] have been suggested and utilized in actual materials.

In this work, we focused on an antiperovskite-type calcium oxide, $Ca_3AO$ (A = Pb, Sn), as a new candidate thermoelectric material with both large $S$ and low $\rho$. $Ca_3PbO$ and $Ca_3SnO$ were first synthesized by Widera and Schäfer in 1980.[10] Both compounds are reported to crystallize in the cubic antiperovskite-type structure with the space group of $Pm\bar{3}m$ between 50 and 500 K on the basis of a single-crystal X-ray diffraction (XRD) analysis.[10,11] Recently, these compounds have been of considerable interest as a candidate Dirac fermion system, as proven by first principles calculations using the generalized gradient approximation (GGA).[12,13,14]

We noticed that the electronic structures of $Ca_3PbO$ and $Ca_3SnO$ could allow them to realize high thermoelectric performance. According to the first principles calculations using GGA, the Pb/Sn 6*p*/5*p* valence band and the Ca 3*d* conduction band are converted at around the Γ point in *k*-space, resulting in them crossing at a certain point on the Γ–X line at the Fermi energy $E_F$. Since the total overlapping of the Pb/Sn 6*p*/5*p* and Ca 3*d* orbitals is zero at this point, because of the requirement of crystal symmetry, a Dirac point emerges. Actually, small band gaps of 14 and 4 meV for the Pb and Sn compounds, respectively, open at the Dirac point due to the strong spin-orbit interaction of the Pb and Sn atoms.[14] However, we must keep in mind that it is sometimes difficult to accurately predict the details of the Fermi surface in the candidate Dirac fermion systems, because their band structure near $E_F$ is often complex. The electronic structure calculations employing Heyd-Scuseria-Ernzerhof screened Coulomb hybrid density functionals (HSE) have recently reported that $Ca_3SnO$ is a normal band insulator with a direct gap of 0.2 eV at the Γ point.[15]

When the band structure reported in the former calculations is realized, small band gaps open at six equivalent points in the first Brillouin zone, suggesting that the $S$ of these compounds can be enhanced by a six-fold degeneracy of the band gap. Moreover, a linear dispersion of the electron energy near the Dirac point can give an extremely high carrier mobility, such as in α-$(BEDT\text{-}TTF)_2I_3$ and $Cd_3As_2$,[16,17] which yield very low ρ. Thus, $Ca_3PbO$ and $Ca_3SnO$ are expected to show both large

a) Electronic mail: yokamoto@nuap.nagoya-u.ac.jp

$S$ and low $\rho$, but there is currently no experimental data on the physical properties of these compounds, except for a magnetoresistance measurement of a single crystal.[18] In this paper, we report the thermoelectric properties of polycrystalline samples of $Ca_3AO$ (A = Pb, Sn). We also tried to dope electron carriers via Bi substitution for A atoms. $Ca_3SnO$ shows metallic $\rho$ and large and positive $S$, approaching 100 $\mu V\ K^{-1}$ at room temperature, suggesting that high potential of antiperovskite calcium oxides as a thermoelectric material.

## II. EXPERIMENTAL

Polycrystalline samples of $Ca_3PbO$ and $Ca_3SnO$ were prepared by solid state reactions with reference to the synthesis of $Ca_3GeO$ and $Ca_3SiO$, as reported by Velden and Jansen.[19] A 4.2 : 1 molar ratio of Ca chips and PbO powder for $Ca_3PbO$ and an 8.4 : 1 : 1 molar ratio of Ca chips, Sn powder, and $SnO_2$ powder for $Ca_3SnO$ were sealed in evacuated quartz tubes. The tubes were heated, kept at 850 °C for 3–6 h, and then quenched. The obtained samples were pulverized, mixed, and then pressed into pellets. These pellets were wrapped with a tantalum foil and sealed in evacuated quartz tubes. The tubes were heated, kept at 850 °C for 1 h, and then quenched. Bi-substituted samples, with a nominal composition of $Ca_3Pb_{1-x}Bi_xO$ ($x$ = 0.1, 0.2), were prepared by applying the above procedure to a 12.6 : 3 − 3$x$ : $x$ : $x$ molar ratio of Ca chips, PbO, Bi, and $Bi_2O_3$ powders. Bi or Sb substitution for the Sn atoms in $Ca_3SnO$ was not successful. The obtained samples were handled in an inert atmosphere because they are air sensitive.

Sample characterization was performed by powder XRD analysis with Cu K$\alpha$ radiation at room temperature using a RINT-2100 diffractometer (RIGAKU) and energy dispersive X-ray spectroscopy (EDX) using Genesis-2000K (EDAX). Electrical resistivity, Hall resistivity, and heat capacity measurements were performed using a physical property measurement system (Quantum Design). Thermoelectric power and thermal conductivity were measured by a standard four-contact method between 100 and 300 K.

## III. RESULTS AND DISCUSSION

### A. Sample Characterization

Figure 1(a) shows powder XRD patterns of $Ca_3Pb_{1-x}Bi_xO$ ($x$ = 0, 0.1, 0.2) and $Ca_3SnO$ polycrystalline samples taken at room temperature. All diffraction peaks, except for some small peaks of tiny amounts of Pb, $Ca_5Pb_3$, and $Ca_4Bi_2O$ impurities, can be indexed on the basis of a cubic structure with a lattice constant of $a$ ~ 4.8 Å, indicating that the cubic antiperovskite phase is successfully obtained as the main phase in all samples. The sharp diffraction peaks indicate good crystallinity of the main phase. The lattice constants $a$ of $Ca_3Pb_{1-x}Bi_xO$ ($x$ = 0, 0.1, 0.2) and $Ca_3SnO$, determined by means of powder XRD, are shown in Fig. 1(b). The $a$ values of $Ca_3PbO$ and $Ca_3SnO$ in the present study are 0.4%–0.5% and 0.04%–0.2% smaller than those of the previous studies,[10,11] respectively, suggesting the presence of a small difference in the chemical compositions between the present and previous studies due to the formation of some defects. No vacancies or intersite defects were observed in the single crystals of $Ca_3PbO$ and $Ca_3SnO$ in the previous study.[11] Due to the difference in synthesis conditions, there may be a marginal amount of vacant sites in the polycrystalline samples, resulting in the slightly smaller $a$ values than those for the single crystals.

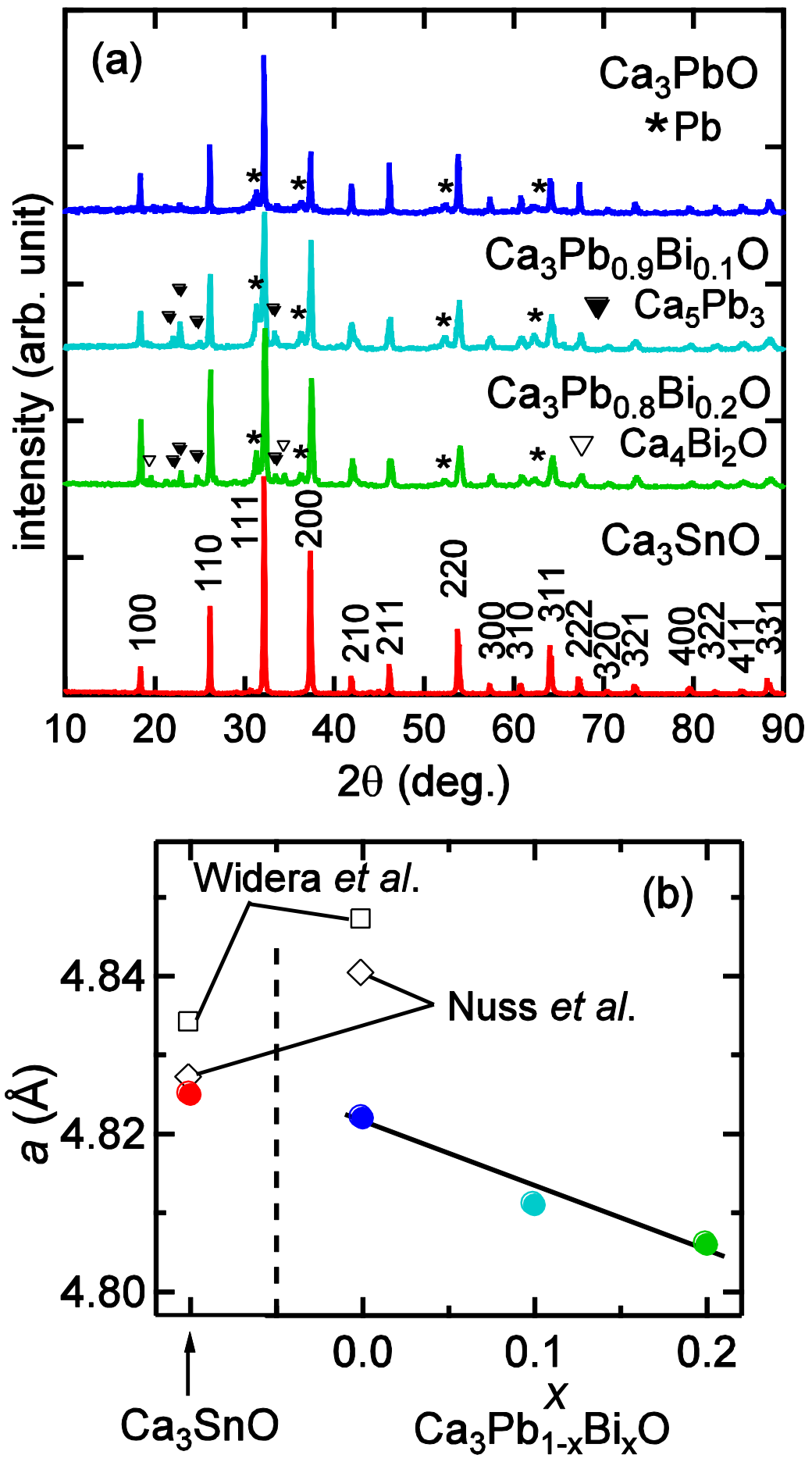

FIG. 1. (a) Powder XRD patterns of $Ca_3Pb_{1-x}Bi_xO$ ($x$ = 0, 0.1, 0.2) and $Ca_3SnO$ polycrystalline samples taken at room temperature. The peaks indicated by stars, filled triangles, and open triangles are those of Pb, $Ca_5Pb_3$, and $Ca_4Bi_2O$ impurities, respectively. Peak indices on the $Ca_3SnO$ data are given using a cubic unit cell with $a$ = 4.8245(8) Å. (b) Lattice constant of $Ca_3Pb_{1-x}Bi_xO$ ($x$ = 0, 0.1, 0.2) and $Ca_3SnO$ polycrystalline samples determined by powder XRD (filled circles). Values reported in the previous studies are indicated by open symbols.[10,11]

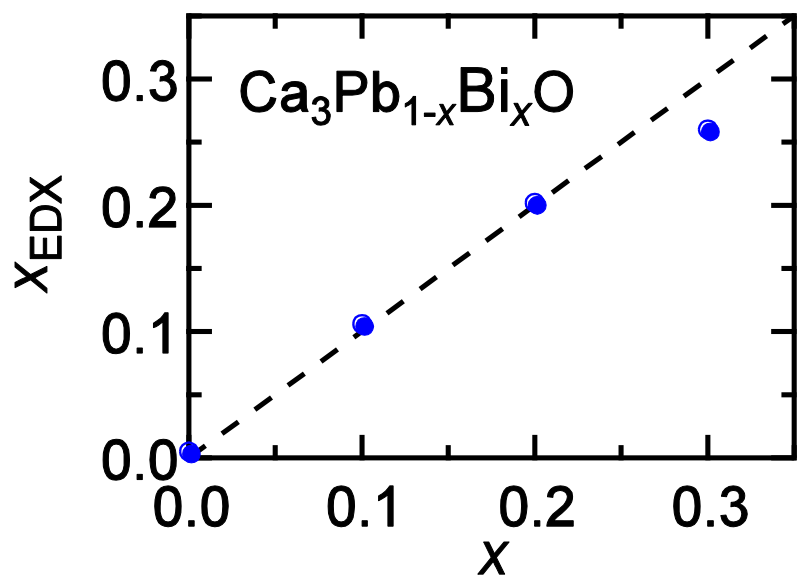

FIG. 2. Bi content estimated by EDX, $x_{EDX}$, as a function of the nominal Bi content $x$ for $Ca_3Pb_{1-x}Bi_xO$ ($x$ = 0, 0.1, 0.2, 0.3) polycrystalline samples. The $x_{EDX} = x$ line is shown by a broken line.

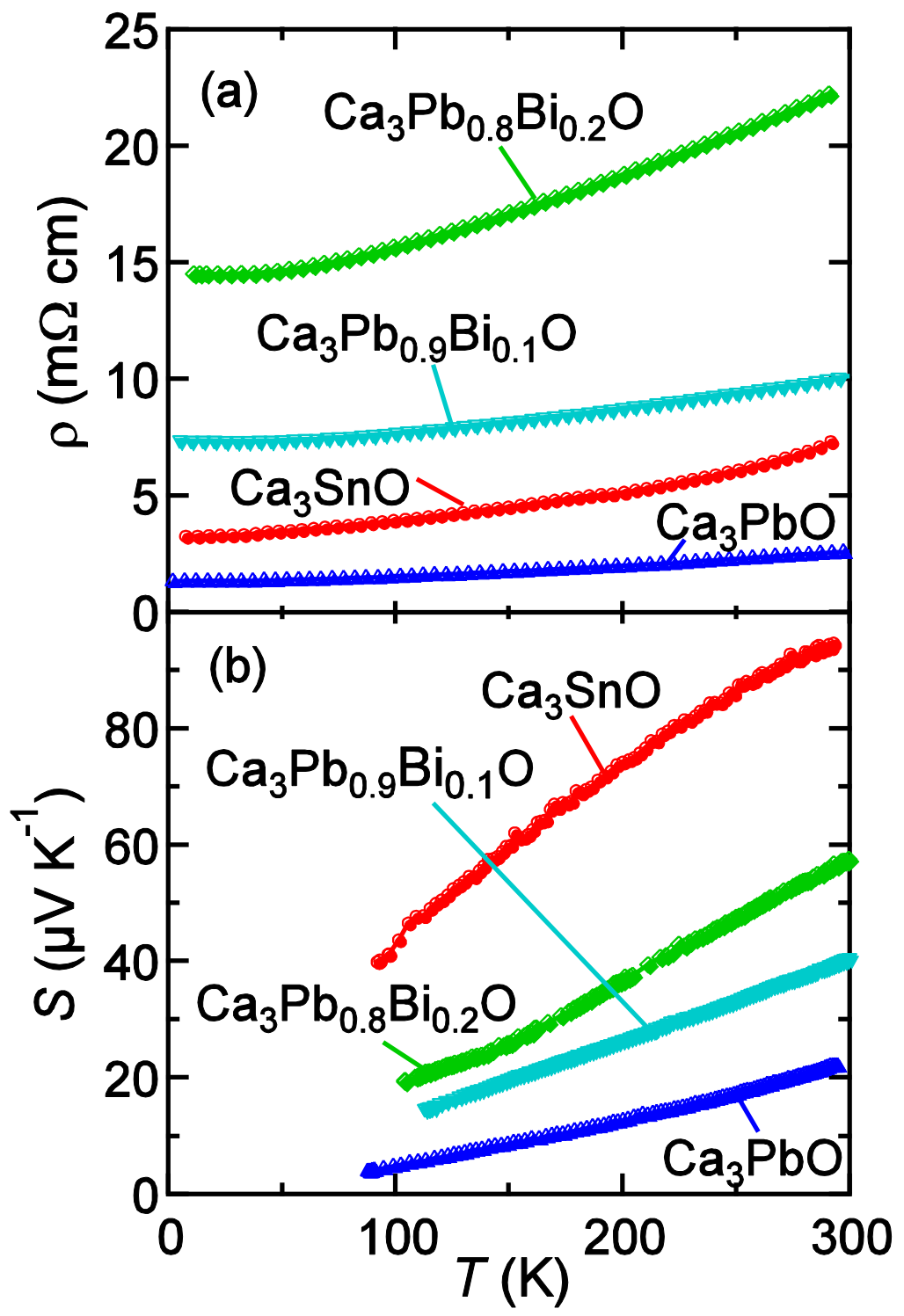

FIG. 3. Temperature dependence of (a) electrical resistivity and (b) thermoelectric power of $Ca_3Pb_{1-x}Bi_xO$ ($x$ = 0, 0.1, 0.2) and $Ca_3SnO$ polycrystalline samples.

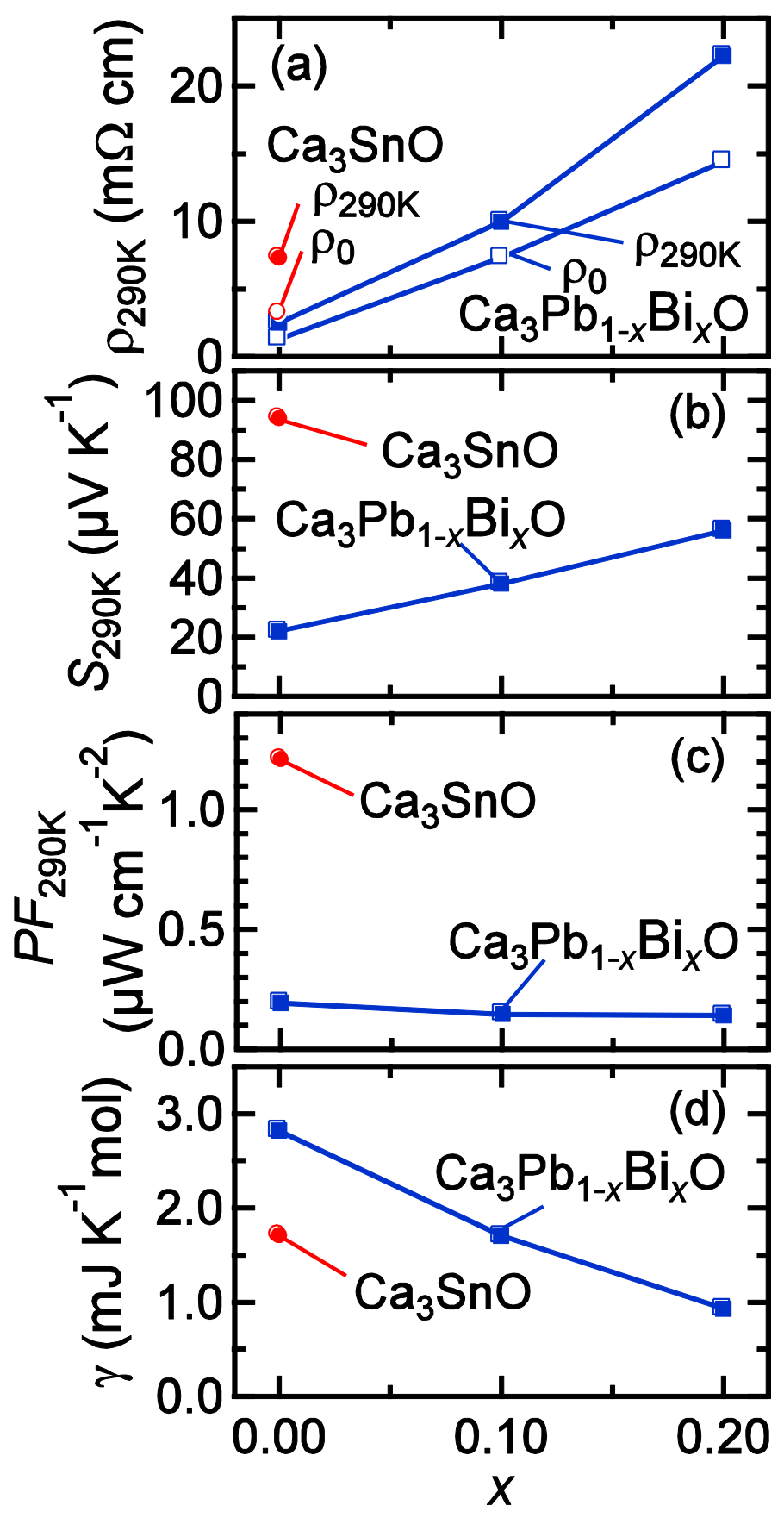

FIG. 4. (a) Electrical resistivity at 290 K, residual resistivity, (b) thermoelectric power at 290 K, (c) power factor at 290 K, and (d) Sommerfeld coefficient of $Ca_3Pb_{1-x}Bi_xO$ ($x$ = 0, 0.1, 0.2) (filled and open squares) and $Ca_3SnO$ (filled and open circles) polycrystalline samples.

As seen in Fig. 1(b), $a$ in $Ca_3Pb_{1-x}Bi_xO$ systematically decreases with increasing $x$, suggesting that the solid solution samples are successfully obtained. The molar ratio of Pb and Bi in the $x$ = 0.1 and 0.2 samples, estimated by EDX, is almost identical to the nominal value, as shown in Fig. 2. In contrast, the estimated molar ratio of Pb and Bi in the $x$ = 0.3 sample is 0.74 : 0.26, different from the nominal ratio of 0.70 : 0.30. In addition, the amount of $Ca_4Bi_2O$ impurity in the powder XRD pattern for $x$ = 0.3 is much larger than for $x \leq 0.2$. These results indicate that the solubility limit of Bi is located at around $x$ = 0.26.

## B. Thermoelectric Properties

Figure 3(a) shows the temperature dependence of the electrical resistivity ρ of the $Ca_3Pb_{1-x}Bi_xO$ ($x$ = 0, 0.1, 0.2) and $Ca_3SnO$ polycrystalline samples. All samples show metallic behavior, where ρ decreases with decreasing temperature. The electrical resistivity of $Ca_3PbO$ is smallest in all samples, electrical resistivity at 290 K $\rho_{290K}$ = 2.5 mΩcm and residual resistivity $\rho_0$ = 1.3 mΩcm. Bi doping into $Ca_3PbO$ increases $\rho_{290K}$ and $\rho_0$; $\rho_{290K}$ = 22.2 mΩcm and $\rho_0$ = 14.4 mΩcm for $x$ = 0.2 are an order of magnitude larger than those for $x$ = 0. The increase of $\rho_0$ may be caused by disorder due to the formation of the solid solution and precipitation of insulating impurities, such as $Ca_4Bi_2O$,[20] at the grain boundaries. Alternatively, $\rho_{290K}$ and $\rho_0$ of the $Ca_3SnO$ sample are 7.3 and 3.2 mΩcm, respectively, which are more than twice those of the $Ca_3PbO$ sample.

The thermoelectric power $S$ of the $Ca_3Pb_{1-x}Bi_xO$ ($x$ = 0, 0.1, 0.2) and $Ca_3SnO$ polycrystalline samples is positive and decreases with decreasing temperature in the whole measured temperature range, as shown in Fig. 3(b). $S$ seems to go to zero toward $T$ = 0 K. These results indicate that these samples are $p$-type metals, where hole carriers are dominant, consistent with the metallic ρ shown in Fig. 3(a). $Ca_3PbO$ shows a small $S$ of 22 μV $K^{-1}$ at 290 K, which increases with increasing Bi composition $x$, as shown in Fig. 4(b). In contrast, the $Ca_3SnO$ sample exhibits a much larger $S$ of 94 μV $K^{-1}$ at 290 K, compared to the $Ca_3Pb_{1-x}Bi_xO$ samples. This $S$ is considerably large for a metallic system, comparable to those of metallic thermoelectric oxides like $Na_xCoO_2$ and $Sr_{1-x}La_xTiO_3$.[21,22] As a result of the large $S$ and metallic ρ, the thermoelectric power factor, $PF = S^2/\rho$, of the $Ca_3SnO$ sample is 1.2 μW $cm^{-1}$ $K^{-2}$ at 290 K, which is an order of magnitude larger than the 0.14–0.19 μW $cm^{-1}$ $K^{-2}$ of the $Ca_3Pb_{1-x}Bi_xO$ samples, indicating that $Ca_3SnO$ is more promising as a thermoelectric material than $Ca_3PbO$.

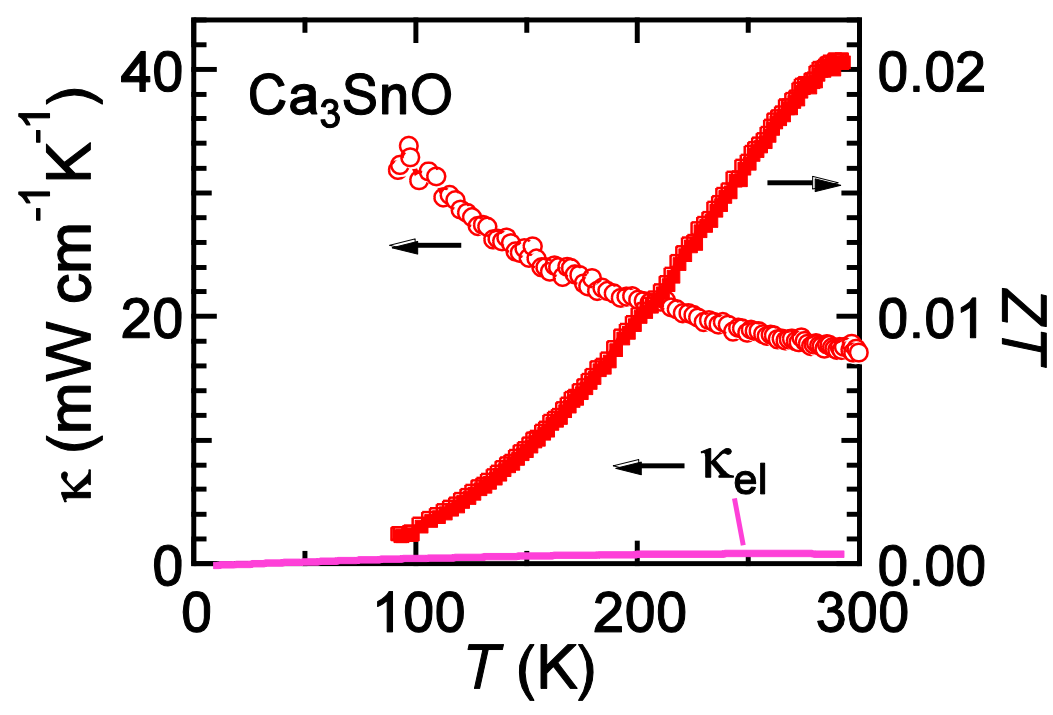

FIG. 5. Temperature dependence of thermal conductivity (left) and dimensionless figure of merit (right) of the $Ca_3SnO$ polycrystalline sample. A solid curve shows the thermal conductivity of conduction electrons (left), estimated by applying the Wiedemann-Franz law to the electrical resistivity data shown in Fig. 3(a).

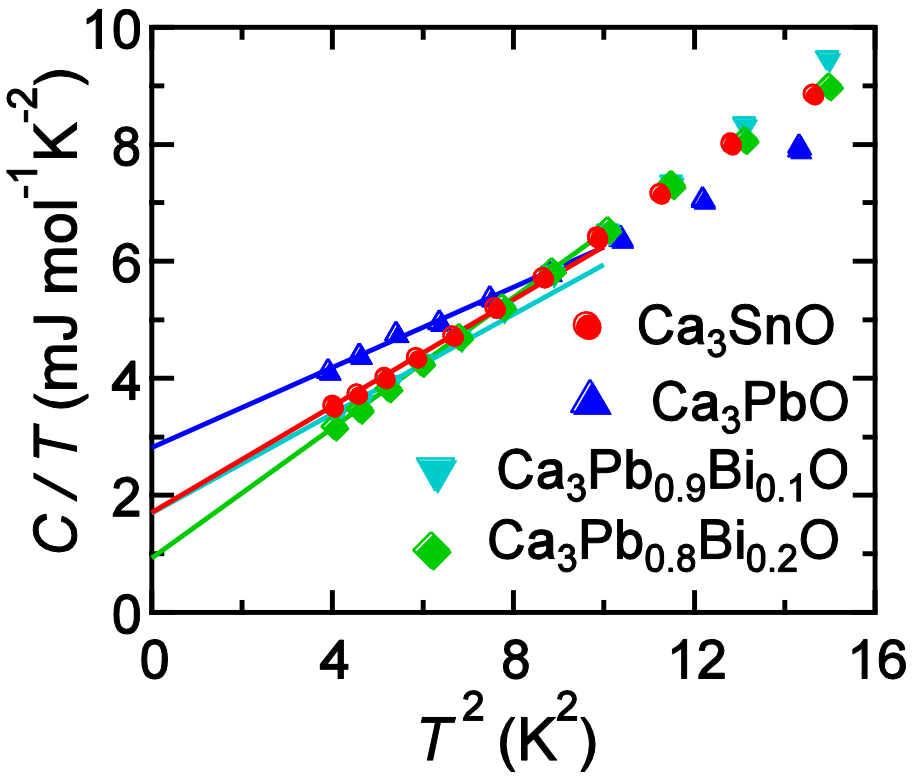

FIG. 6. Heat capacity divided by temperature of $Ca_3Pb_{1-x}Bi_xO$ ($x$ = 0, 0.1, 0.2) and $Ca_3SnO$ polycrystalline samples as a function of $T^2$. The solid lines show results of the linear fits between 2 and 3 K for each sample.

As shown in Fig. 5, the thermal conductivity κ of the $Ca_3SnO$ polycrystalline sample decreases with increasing temperature and becomes 17 mW $cm^{-1}$ $K^{-1}$ at 290 K. The thermal conductivity of conduction electrons $\kappa_{el}$ is estimated to be less than 1 mW $cm^{-1}$ $K^{-1}$, as shown in Fig. 5, by applying the Wiedemann-Franz law, $\kappa\rho = LT$, where $L$ is Lorentz number of 2.44 × $10^{-2}$ mW mΩ $K^{-2}$, to the electrical resistivity data. This estimation indicates that the lattice contribution is dominant in the κ of the $Ca_3SnO$ sample, consistent with the $T^{-1}$ temperature dependence above 100 K. The value of κ = 17 mW $cm^{-1}$ $K^{-1}$ at 290 K is larger than those of glasses and amorphous alloys (κ = 5–10 mW $cm^{-1}$ $K^{-1}$), but relatively small for crystalline inorganic compounds.

The dimensionless figure of merit $ZT$ of the $Ca_3SnO$ sample is shown in Fig. 5. $ZT$ increases with increasing temperature and reaches 0.02 at 290 K. This value is much smaller than those of current thermoelectric materials for practical applications. However, almost half of ρ at room temperature is the contribution of residual resistivity mostly caused by an extrinsic origin, which can be reduced by improving the electrical conductivity of the grain boundary with little effect on $S$. Moreover, suppressing lattice thermal conductivity $\kappa_{lat}$ to be as low as an amorphous value, maintaining the good electrical conductivity by forming the solid solution with $Sr_3SnO$ or $Ba_3SnO$ will significantly improve $ZT$.

## C. Electronic State

We now discuss the relationship between the electronic state and thermoelectric properties of $Ca_3AO$. As seen in Fig. 3, the $Ca_3Pb_{1-x}Bi_xO$ ($x$ = 0, 0.1, 0.2) and $Ca_3SnO$ samples show metallic ρ and positive $S$ for the entire measured temperature range, suggesting that the Fermi energy $E_F$ is located just below the top of the valence band. Figure 6 shows the heat capacity $C$ divided by the temperature of the $Ca_3Pb_{1-x}Bi_xO$ ($x$ = 0, 0.1, 0.2) and $Ca_3SnO$ polycrystalline samples as a function of $T^2$. Linear fits to the low temperature part of the data yield finite Sommerfeld coefficients of γ = 1–3 mJ $K^{-2}$ $mol^{-1}$ for all samples. The γ of $Ca_3Pb_{1-x}Bi_xO$ decreases with increasing $x$, meaning the electronic density of states at the Fermi energy $D(E_F)$ decreases with Bi substitution, i.e., electron doping, supporting the suggestion that the $E_F$ lies just below the top of the valence band. The increase of $S$ with Bi doping as seen in Fig. 3(b) is also consistent with this idea. The stoichiometric $Ca_3PbO$ and $Ca_3SnO$ should show insulating behavior at low temperature, because the $E_F$ is located in the small band gap predicted by the band calculation in that case.[12,14] The polycrystalline samples in the present study are suggested to be significantly hole-doped by some defects, resulting in metallic behavior in physical properties.

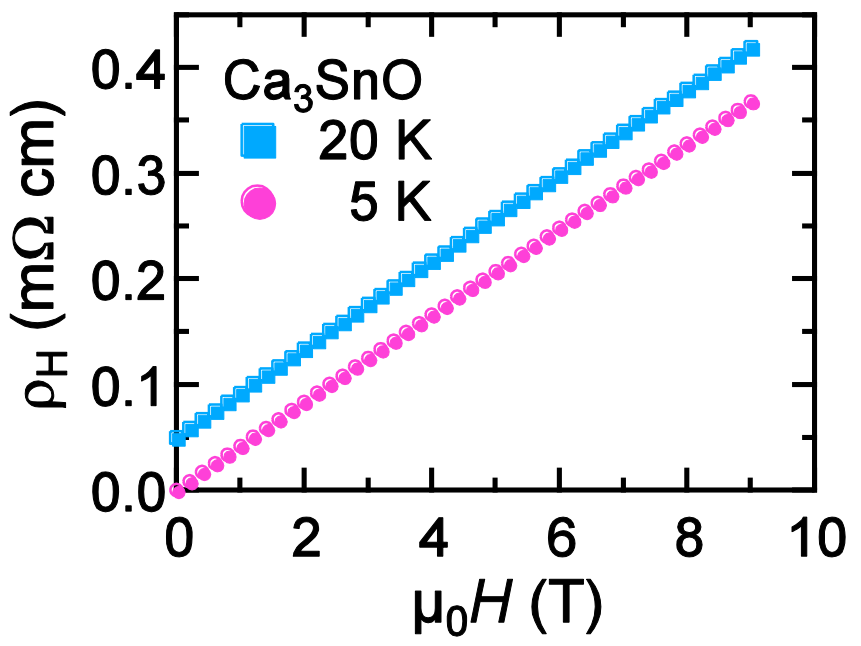

FIG. 7. Hall resistivity of the $Ca_3SnO$ polycrystalline sample measured in magnetic fields up to 9 T taken at 5 and 20 K. The 20 K data are shifted by 0.05 mΩ cm for clarity.

Comparing the $Ca_3Pb_{1-x}Bi_xO$ ($x$ = 0, 0.1, 0.2) and $Ca_3SnO$ samples from the view point of thermoelectricity, $Ca_3SnO$ shows a higher performance than the Pb compounds. In particular, the $Ca_3SnO$ sample shows more than twice the value of $S$ than that of the $Ca_3Pb_{0.9}Bi_{0.1}O$ sample, despite the $\gamma$ of them being almost the same, indicating that they have similar $D(E_F)$, but the $|dD(E_F)/dE|$ of the former is much larger than that of the latter. Hence, the $Ca_3SnO$ sample shows similar $\rho$ and larger $S$ compared to $Ca_3Pb_{0.9}Bi_{0.1}O$, resulting in larger $PF$ in the $Ca_3SnO$ sample than those of the Pb samples, as seen in Fig. 4(c).

Here, we consider the high thermoelectric performance of $Ca_3SnO$ from the properties of the conducting carriers. To realize high thermoelectric performance, conducting carriers should simultaneously have high mobility, large effective mass, and high degeneracy of the band occupied by them. The former gives rise to low $\rho$, while the latter two result in large $S$. In fact, $Ca_3SnO$ meets these requirements. The Hall resistivity $\rho_H$ of the $Ca_3SnO$ sample, measured to estimate the mobility and carrier density, is shown in Fig. 7. The $\rho_H$ linearly increases with increasing a magnetic field. Linear fits of the data yield Hall coefficients $R_H$ of 0.435 and 0.437 $cm^3$ $C^{-1}$ and carrier densities of $n = R_H/e$ of 1.44 × $10^{19}$ and 1.43 × $10^{19}$ $cm^{-3}$ at 5 and 20 K, respectively, where $-e$ is the electron charge. This positive $R_H$ indicates that hole carriers are dominant in the $Ca_3SnO$ sample, consistent with the positive thermoelectric power. On the other hand, this $n$ is significantly low as a metallic system. The mobilities, $\mu = R_H/\rho$, are estimated to be 79.4 and 78.4 $cm^2$ $V^{-1}s^{-1}$ at 5 and 20 K, respectively, which are smaller than those of high-mobility semiconductors, such as practical thermoelectric materials $Bi_2Te_3$ and PbTe,[1] but much larger than those of normal metals. This relatively high $\mu$ for a metallic system is an important factor which gives rise to the high thermoelectric performance in $Ca_3SnO$.

An interesting point regarding the thermoelectric properties of $Ca_3SnO$ is the large $\gamma$ for a low-carrier metal with $n \sim 10^{19}$ $cm^{-3}$, which results in the coexistence of high mobility and a large effective mass. Assuming that the six hole pockets exist in the first Brillouin zone, according to the first principles calculation using GGA,[12,14] and the top of the valence band is parabolic, the effective mass of the hole carrier $m^*$ is estimated to be 1.96$m$ from $\gamma$ = 1.71 mJ $K^{-2}$ $mol^{-1}$ and $n$ = 1.44 × $10^{19}$ $cm^{-3}$, where $m$ is the mass of a bare electron. This $m^*$ is comparable to that of a normal metal with $10^{22}$–$10^{23}$ $cm^{-3}$ and considerably larger than semiconducting thermoelectric materials with $n \sim 10^{19}$ $cm^{-3}$, such as $Bi_2Te_3$ and PbTe ($m^* \sim 0.2m$ for $n$-type $Bi_2Te_3$).[1] Thus, $Ca_3SnO$ simultaneously has higher $\mu$ and larger $m^*$ than those of normal metals and semiconductors, respectively, and high degeneracy of the valence band of $N$ = 6, according to the first principles calculations using GGA, leading to the high thermoelectric performance in this compound.

The thermoelectric B factor,[1,23] $B = 2N\mu k_B{}^2T(2\pi m^* k_B T/h^2)^{3/2}/e\kappa_{lat}$, of $Ca_3SnO$ is estimated to be $B$ = 0.68 at 290 K, where $k_B$ and $h$ are the Boltzmann and Planck constants, respectively. The B factor is a dimensionless number to estimate the efficiency of a thermoelectric material as a function of $\mu$, $m^*$, $N$, and $\kappa_{lat}$. These four material properties can be controlled independently different from $S$, $\rho$, and $\kappa$ in $ZT$. The B factor gives the maximum $ZT$ when the size of a band gap and $n$ are optimized, i.e., a material with larger $B$ has a larger upper limit of $ZT$. A value of $B$ = 0.68 in $Ca_3SnO$ is larger than those of practical thermoelectric materials $Bi_2Te_3$ and PbTe ($B$ = 0.3-0.4) and indicates the upper limit of $ZT \sim 1.9$. $B$ is not necessarily linked to $ZT$ of an obtained sample, but this large $B$ suggests the excellent potential of $Ca_3SnO$ as a thermoelectric material. On the other hand, if we apply above discussion to the case of the electronic structure calculations employing HSE,[15] where the valence band is doubly degenerated ($N$ = 2), $m^*$ and $B$ are estimated to be 1.36$m$ and 0.13, respectively. These $m^*$ and $B$ are moderately large and the $B$ gives an upper limit of $ZT \sim 0.6$, but significantly smaller than that in the $N$ = 6 case.

## IV. CONCLUSIONS

$Ca_3Pb_{1-x}Bi_xO$ ($x$ = 0, 0.1, 0.2) and $Ca_3SnO$ polycrystalline samples are found to show metallic $\rho$ and positive $S$ in the whole measured temperature range, indicating $E_F$ in these samples are located just below the top of the valence band. The $S$ of the $Ca_3SnO$ sample approaches 100 μV $K^{-1}$ at room temperature, which is much larger than those of the Pb samples, resulting in larger $PF$ in the $Ca_3SnO$ sample. The high thermoelectric performance of $Ca_3SnO$ is mostly caused by (i) the coexistence of moderately high $\mu$ and large $m^*$ and (ii) the possible presence of a multi-valley electronic structure with six small hole pockets in the first Brillouin zone.

## ACKNOWLEDGMENTS

This work was partly carried out at the Materials Design and Characterization Laboratory under the Visiting Researcher Program of the Institute for Solid State Physics, Univ. of Tokyo and partly supported by JSPS KAKENHI (Grant Number: 25800188, 16H03848, and 16K13664), Research Foundation for the Electrotechnology of Chubu, and Iketani Science and Technology Foundation.